\documentclass[12pt,letterpaper]{article}
\pdfoutput=1
\usepackage{psk_v1}
\usepackage{threeparttable}
\newcommand{\text}[1]{\mathrm{#1}}

\begin{document}

\title{Low-power continuous-wave four-wave mixing wavelength conversion in AlGaAs-nanowaveguide microresonators}

\author{Pisek Kultavewuti,$^{1,\ast}$ Vincenzo Pusino,$^2$ Marc Sorel,$^2$ and \mbox{J. Stewart Aitchison$^1$}}

\address{$^1$Department of Electrical and Computer Engineering, University of Toronto, 10 King's College Road, Toronto, Ontario, Canada M5S 3G4\\
$^2$School of Engineering, University of Glasgow, Glasgow G12 8QQ, Scotland, United Kingdom}

\email{$^*$pisek.kultavewuti@mail.utoronto.ca} %% email address is required



\begin{abstract*}
We experimentally demonstrated enhanced wavelength conversion in a Q$\sim$7500 deeply etched AlGaAs-nanowaveguide microresonator via degenerate continuous-wave four-wave mixing with a pump power of $24\;\mathrm{ mW}$. The maximum conversion efficiency is $-43\;\mathrm{ dB}$ and accounts for $12\;\mathrm{ dB}$ enhancement compared to that of a straight nanowaveguide. The experimental results and theoretical predictions agree very well and show optimized conversion efficiency of $-15\;\mathrm{ dB}$. This work represents a step toward realizing a fully integrated optical devices for generating new optical frequencies.
\end{abstract*}

%\ocis{(270.0270) Quantum optics; (190.4390) Nonlinear optics, integrated optics; (190.4380) Nonlinear optics, four-wave mixing.}
%For a complete list of OCIS codes, visit: http://www.opticsinfobase.org/submit/ocis/

%%%%%%%%%%%%%%%%%%%%%%%%%%  body  %%%%%%%%%%%%%%%%%%%%%%%%%%
\linespread{1.3}\selectfont
\vskip0.5in
Nonlinear integrated optical devices hold a lot of promises for the generation of new frequencies and entangled photons. Nanoscale semiconductor waveguiding structures have several advantages over bulk crystals or fibers. Nanowaveguides concentrate optical powers to boost optical intensities, effectively reducing the power requirement for efficient nonlinear interactions, in particular wavelength conversion. They also have potential for integration leading to multiple functionalities in a single device and even hybridization with electronics to utilize strengths of both technologies.

Four-wave mixing (FWM) has been extensively studied and has the capacity of implementing a broad range of functionalities. For optical signal processing, FWM can provide signal regeneration \cite{Salem2007}, wavelength channel conversion \cite{Dolgaleva:2011wb}, and logic operations \cite{Li2011}, all of which could be modulation-format transparent and compatible with coherent communications \cite{Adams2014}. FWM-based wavelength conversion can help in translating signals in a wavelength range, such as mid-IR, that is difficult to detect to another range that has very efficient detectors \cite{Liu:2012tp}. When FWM is highly efficient, a signal wave can be amplified and possibly oscillated in a proper cavity, leading to a widely tunable light source covering regions unreachable by conventional laser media \cite{Okawachi2011}. In fact, while implementing FWM in a microresonator, FWM can generate a very broadband frequency comb, which is a cornerstone in many applications such as metrology, precise clocking, and spectrocopy.

Many materials have been proposed for integrated FWM-based wavelength conversion devices including crystalline silicon \cite{Turner-Foster2010,Kuyken2014}, silicon nitride \cite{Lamont2013}, chalcogenide glass \cite{Luan2009,Ahmad2012}, Hydex \cite{Ferrera2008}, and AlGaAs \cite{Dolgaleva:2011wb, Mahmood2014, Lacava:2014wk}. Silicon has high nonlinear refractive index $n_2$ but suffers from two-photon absorption (2PA) and free-carrier absorption (FCA) in the C-band despite low linear propagation losses. Silicon nitride and Hydex greatly mitigate these 2PA and FCA adversities but they lack potential integration with active devices. Chalcogenide has considerably lower $n_2$ and its fabrication process is rather sophisticated. On the other hand, AlGaAs possesses high nonlinear refraction and offers bandgap and refractive index engineering via aluminum molar concentration. It has low 2PA and FCA effects in the C-band due to operation below a half bandgap \cite{Aitchison1997,Wathen2014}. It also has capacity for implementing active and passive components and hybridization with other III-V semiconductors.

There are only a few studies on FWM in passive AlGaAs nanowaveguides in the past primarily because of high propagation losses from sidewall scattering \cite{Siviloglou2006}. Therefore, pulsed pumps had to be used to perform the mixing. Nonetheless, broadband FWM in high-loss AlGaAs nanowaveguides in the anomalous dispersion regime has been reported \cite{Dolgaleva2013}. Recent fabrication techniques produce low-loss nanowaveguides with loss figures from 0.56 to $5 \;\text{dB/cm}$ \cite{Apiratikul2014a, Lacava:2014wk}. Additionally, the nanoscale dimension of the waveguides boosts the photon-photon interaction, nominally increasing the nonlinear coefficient $\gamma$, and critically provides excellent phase matching conditions due to strong waveguide dispersion contribution, yielding a zero-dispersion wavelength (ZDW) in the C-band \cite{Meier2007}. Both of the low loss and the enhanced nonlinear coefficient help in reducing the power requirement for FWM so that continuous-wave (cw) operation can be performed.

A nonlinear interaction can be further enhanced by using a microresonator. AlGaAs microresonators have been investigated extensively in a context of all-optical switching; however, there are fewer reports which focus on FWM-based wavelength conversion. \mbox{Absil \textit{et al.}} reported enhanced FWM due to field enhancement effects from the resonator even with high propagation losses in the device  \cite{Absil2000}. In their work, they employed a pulsed pump with an approximate peak power of $150\;\text{mW}$.

In this paper, we report on the demonstration of continuous-wave FWM wavelength conversion in an \mbox{AlGaAs} microresonator with a maximum conversion efficiency of $-43 \;\text{dB}$ which is enhanced by $12\;\text{dB}$ compared to that of a straight waveguide when pumping at $24\;\text{mW}$.

%%%\section{Devices and Linear Characterization}
\begin{figure}[t]
	\centering
	\includegraphics[width=8cm]{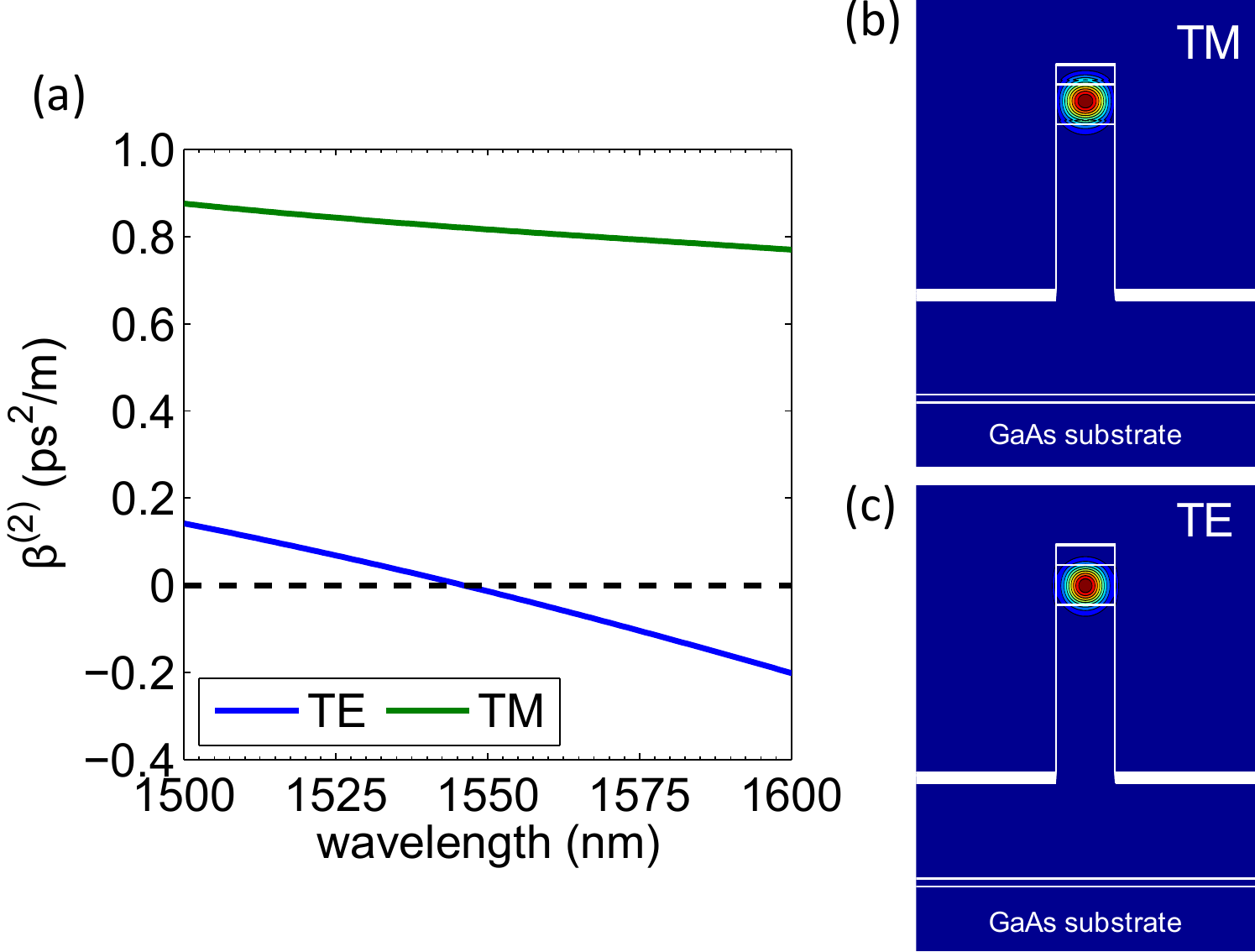}
	\caption{ (a) The second-order dispersion coefficients, $\beta^{(2)}$, for the fundamental TE and TM modes of the 700-nm-wide deeply etched AlGaAs nanowaveguide. (b) and (c) Intensity profiles of the TM and TE modes, respectively, superimposed with framed AlGaAs structures. The modes reside mostly in the core AlGaAs layer with 25\% aluminum. The TE mode has the ZDW near 1550 nm whereas the TM mode is normally dispersive in this wavelength range.}
	\label{fig:dispersion}
\end{figure}
We fabricated devices in a MOVPE AlGaAs structure grown on top of a GaAs substrate. The $\text{Al}_x\text{Ga}_{1-x}\text{As}$ structure consists of layers with aluminum molar concentrations, $x$, of 0.75, 0.25, and 0.75 and with thickness of 4,200, 600, and $300\;\text{nm}$, respectively. The $\text{Al}_{0.25}\text{Ga}_{0.75}\text{As}$ is the guiding core of the waveguide and its aluminum concentration was chosen such that the photon energy at 1550 nm is below the half-bandgap energy, hence mitigating the 2PA effect. The patterns, including straight waveguides and ring-bus systems, were transferred using electron-beam lithography with an HSQ mask and a subsequent dry-etching.

\begin{figure}[t]
	\centering
	\includegraphics[width=6.5cm]{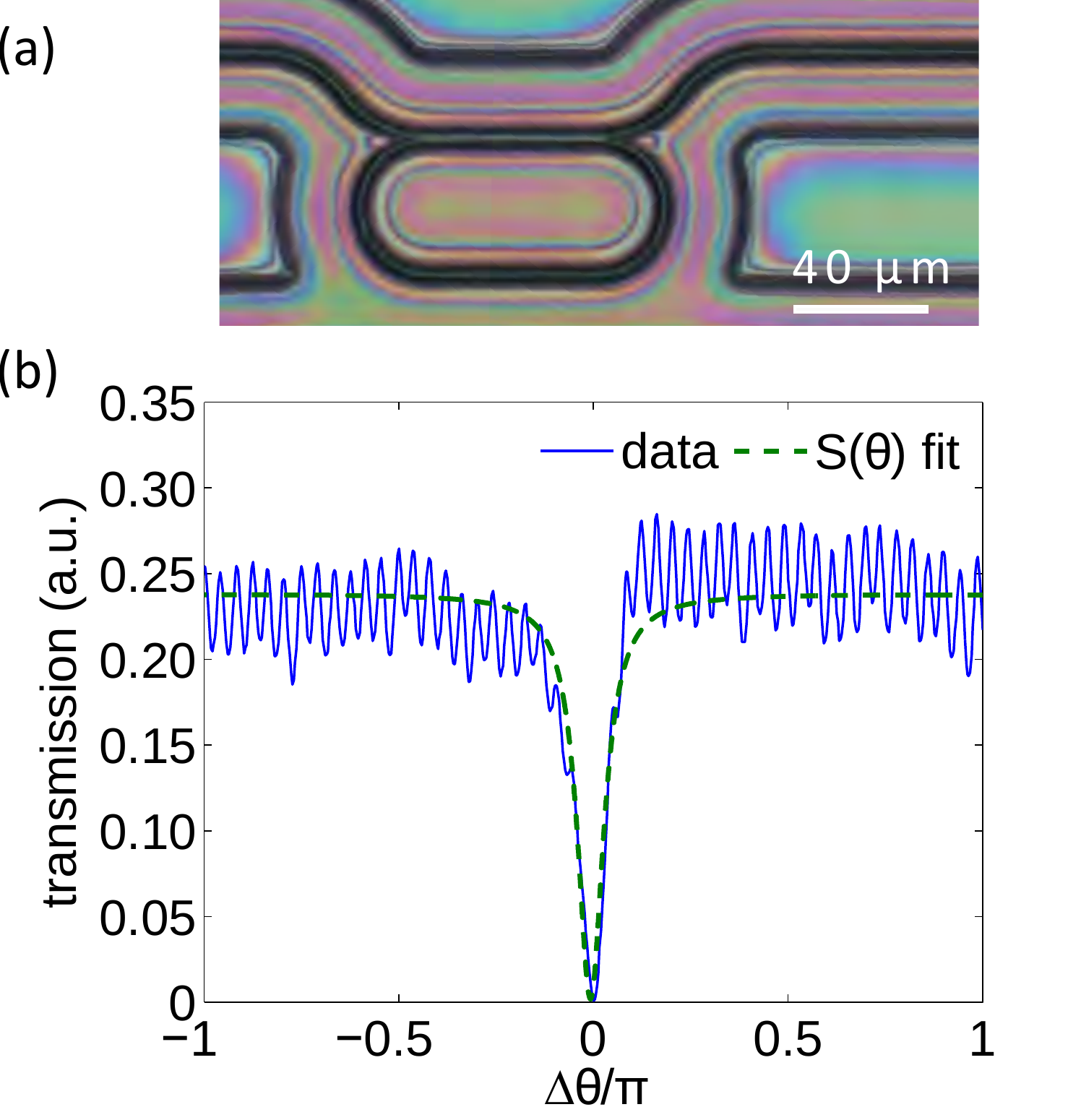}
	\caption{(a) An image of the resonator taken with an optical microscope. (b) Measured transmission of the resonator for the TM excitation along with a fit to the model $S(\theta)$ in Eq. \ref{eq:ringtransmission} versus detuning from the resonance near $1554.9\;\text{nm}$. The resonator exhibits a Q-factor of 7500.}
	\label{fig:ringtx}
\end{figure}

The resultant single-mode waveguides have widths of 700 nm. Using the finite-difference eigenmode solver, Lumerical MODE, we calculate the total dispersion of the waveguide in $\beta^{(2)}=\text{d}^2\beta(\omega)/\text{d}\omega^2$ for the fundamental TE and TM modes and plot the results in Fig. \ref{fig:dispersion}(a), noting that $\beta$ is the propagation constant. Fig. \ref{fig:dispersion}(b) and \ref{fig:dispersion}(c) show spatial profiles of the TM and TE modes, respectively. These nanowaveguides were tapered to $2\text{-}\mu\text{m}$ waveguides for efficient end-fire coupling. The tapers and the $2\text{-}\mu\text{m}$ waveguides account for a combined length of about $0.5\;\text{mm}$. The facet-to-facet straight waveguide length is about $5\;\text{mm}$. For the ring-bus system, the microresonator is actually a race-track of a cavity length of $275.6\;\mu\text{m}$ and it is coupled to a $700\text{-nm-wide}$ bus waveguide at its straight waveguide part. The image of the resonator is shown in Fig. \ref{fig:ringtx}(a).

The propagation loss of the TM mode is $12\;\text{dB/cm}$ as measured from straight waveguides using the Fabry-Perot technique \cite{Walker1985}. The experimentally measured transmission of the resonator for TM excitation is shown in Fig. \ref{fig:ringtx}(b) and fitted to a theoretical transmission \cite{Niehusmann2004}:
\begin{equation}
\label{eq:ringtransmission}
S = \frac{\sigma^2+\tau^2-2\sigma\tau\cos\theta}{1+\sigma^2\tau^2-2\sigma\tau\cos\theta}
\end{equation}
where $\sigma$ is a field transmission coefficient of the ring-bus coupler, $\tau$ is a round-trip field propagation loss, and $\theta$ is a round-trip phase. At the directional coupler, we have $\sigma^2+\kappa^2=1$, where $\kappa$ is a field cross-coupling coefficient. Fitting the measured transmission yields $\sigma=0.94,\kappa=0.34,\tau=0.93$, indicating that our resonator operates near the critical coupling. Note that the off-resonance transmission of the ring is $\sim 0.75$. The resonator has a free spectral range (FSR) of $380\;\text{GHz}$. The superimposing fringe on the measured transmission is attributed to the Fabry-Perot formed by the end facets of the chip. The measured Q-factor of the resonator is $\sim7500$ corresponding to a photon lifetime of about $6.1\;\text{ps}$ and an effective propagation length of $l_p=620\;\mu\text{m}$ before the photon decays or out-couple to the bus waveguide. Comparison between the transmitted powers of the on- and off-resonance cases allows us to estimate the field enhancement inside the resonator, which is given by
\begin{equation}
F = \frac{\kappa}{|1-\sigma\tau e^{-i\theta}|}.
\end{equation}

%%%\section{FWM Wavelength Conversion}
Degenerate-FWM wavelength conversion experiment is conducted as the following. The cw pump and cw signal waves are coupled into the bus using an objective lens and the output optical wave is collected by another objective lens before being split to power meters and an optical spectrum analyzer (OSA). The pump wave comes from a tunable laser, and is amplified by an Erbium-doped fiber amplifier (EDFA). The amplified pump is filtered with a $2\text{-nm-bandpass}$ thin-film fiber filter centering at $1554.9\;\text{nm}$. The signal is derived from another tunable laser without any amplification.

In measuring the FWM spectra and the conversion efficiency (CE), the strong pump is tuned to resonance by soft-thermal locking \cite{Grudinin2009} and the much weaker signal is scanned across on- and off- resonance positions. The coupled pump power is $24\;\text{mW}$ at the input facet, and the circulating power inside the resonator is $6.8\;\text{mW}$ ($8.3\;\text{dBm}$).

\begin{figure}[tb]
	\centering
	\includegraphics[width=8.5cm]{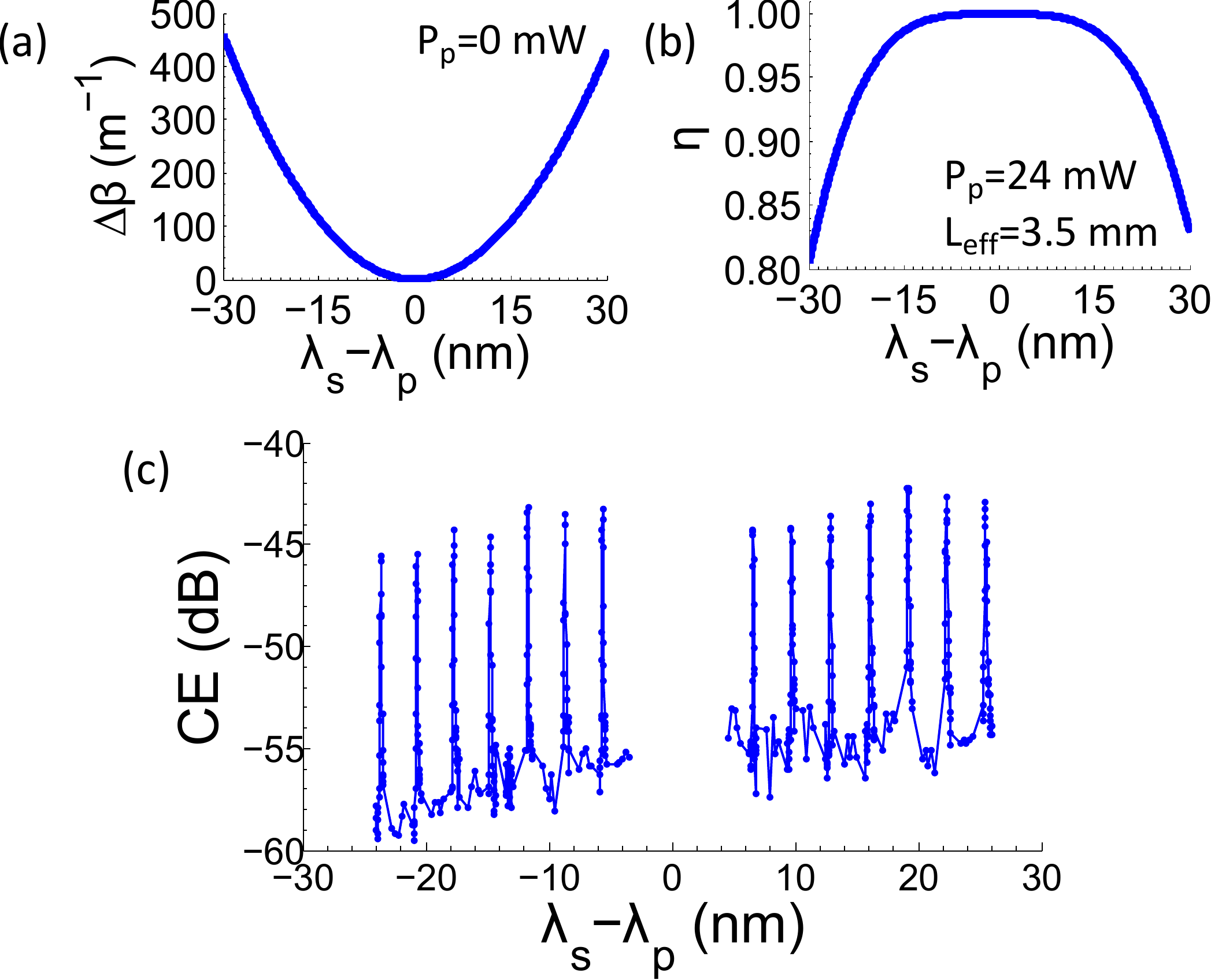}
	\caption{(a) A linear FWM phase matching as a function of signal wavelengths detuning from a pump wavelength at $1555\;\text{nm}$, and (b) the corresponding $\eta$ for a pump power of $24\;\text{mW}$ and an effective length of $3.5\;\text{mm}$ of a 700-nm-wide nanowaveguide. (c) Conversion efficiency versus a signal-pump wavelength detuning where the pump is set to the resonance wavelength around $1555\;\text{nm}$ and the coupled pump power is $24\;\text{mW}$.}
	\label{fig:CE_vs_dlambda}
\end{figure}

The CE is defined as $10\log(P_{i,\text{out}}/P_{s,\text{in}})$, and simplified nonlinear equations can be used to predict the undepleted-pump CE including the resonance effects from \cite{Absil2000}:
\begin{equation}
\label{eq:FWM1}
P_{i,\text{out}} = (\gamma P_p L_\text{eff})^2\cdot P_{s,\text{in}}\cdot e^{-\alpha L}\cdot\eta\cdot F_p^4F_s^2F_i^2,
\end{equation}
\begin{equation}
\label{eq:eta}
\eta = \frac{\alpha^2}{\alpha^2+(\Delta\beta)^2}\Bigg[1+\frac{4e^{-\alpha L}\sin^2(\Delta\beta L/2)}{(1-e^{-\alpha L})^2}\Bigg].
\end{equation}
For the off-resonance cases, the field enhancement factors, $F_x$, are unity. $L_\text{eff}=(1-e^{-\alpha L})/\alpha$ is an effective length calculated from the actual physical length $L$, where FWM interaction occurs. The parameter $\eta$ incorporates the effects of the propagation loss and the total phase mismatch $\Delta\beta=\beta_s+\beta_i-2\beta_p+2\gamma P_p$ on the FWM efficiency, and it informs the conversion efficiency bandwidth. Note that $\eta$ is unity when the phase mismatch is zero. Fig. \ref{fig:CE_vs_dlambda}(a) and \ref{fig:CE_vs_dlambda}(b) plot $\Delta\beta$ and $\eta$ against the signal-pump wavelength detuning for the all-TM FWM in straight waveguides, and the FWM bandwidth is $\sim 80\text {nm}$ as partially displayed.

A preliminary cw-FWM measurement in a straight waveguide suggests that the nonlinear coefficient is $\gamma\approx 150\;\text{m}^{-1}\text{W}^{-1}$. This is achieved by exciting the waveguide with a strong pump and a weak signal whose wavelength is very close to that of the pump. The nonlinear coefficient is then calculated from Eq. \ref{eq:FWM1} with the measured powers, the propagation loss, $\eta=1$, and $F_x=1$.

Considering signal-pump detuning around $-10\;\text{nm}$, the phase mismatch is simulated to be around $|\Delta\beta|=60\;\text{m}^{-1}$. For the off-resonance signal cases, the idler wave is generated mostly in the nanowaveguide section before the resonator which is $1.5\;\text{mm}$ long and leads to the effective length of $L_\text{eff}=1.2\;\text{mm}$ and correspondingly $\eta\approx 1$. This set of parameters yields $\text{CE}=-55\;\text{dB}$. For the on-resonance signal cases, we assume that FWM occurs only in the resonator, justified by 10-times-larger idler powers compared to the off-resonance case. To estimate the CE for this case, we can relate the effective FWM interaction length to the pump photon lifetime in the resonator, i.e. $L_\text{eff}=l_p=620\;\mu\text{m}$, and set the resonator-enhanced pump power $P_pF_p^2$ to the circulating power of $6.8\;\text{mW}$. The field enhancement for the signal, $F_s$, extracted from the measurement is $\sim 3$, and we could assume the same figure for the idler, $F_i\approx F_s$, because of the conservation of energy $\omega_s+\omega_i=2\omega_p$ and the equidistance of the resonator modes within the bandwidth under consideration. This leads to $\text{CE}=-45\;\text{dB}$. Comparing the two cases, the CE enhancement is predicted to be $10\;\text{dB}$, which is in good agreement with the observed value of $12\;\text{dB}$.

\begin{figure}[t]
	\centering
	\includegraphics[width=6.5cm]{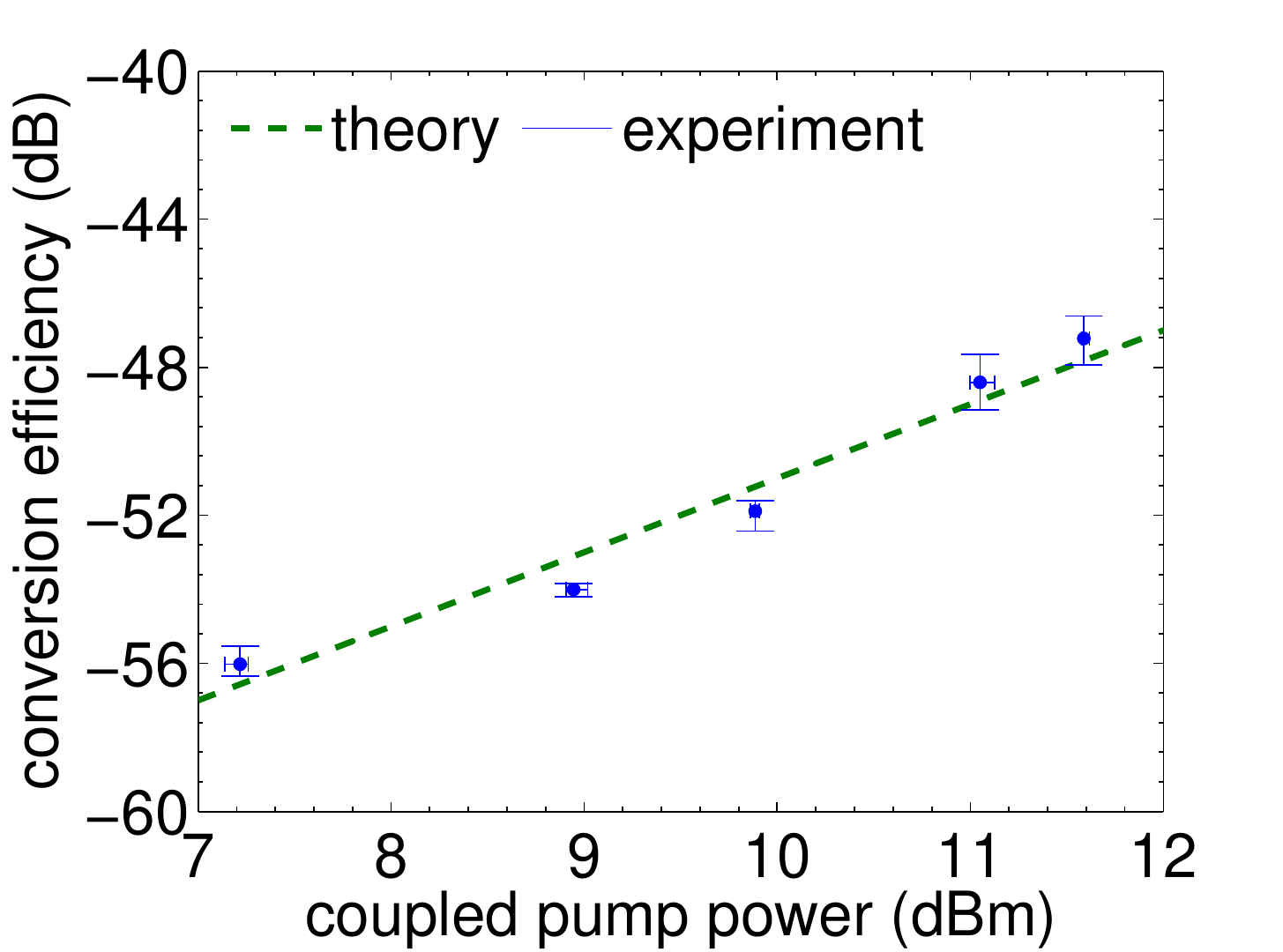}
	\caption{Conversion efficiency as a function of pump powers, in dBm, when the pump wavelength, signal wavelength, and signal power remain constant. The dashed curve is a theoretical prediction based on the simplified nonlinear equation \mbox{Eq. \ref{eq:FWM1}} with a slope of 2.}
	\label{fig:CE_vs_power}
\end{figure}

We estimate the field enhancement for the pump $F_p$ to be around $0.5$\textendash$0.7$ and attribute it to difficulty in keeping the pump tuned to the resonance. This is due to back-reflection from the waveguide facet to the EDFA, which causes power fluctuation. If $F_p=3$, based on Eq. \ref{eq:FWM1} the CE enhancement will be $40\;\text{dB}$ and we would have a peak CE in this experimental setup of $-15\;\text{dB}$.

We also measure CE as a function of pump powers. In this experiment, the pump power is varied from high to low, along with necessary change in its wavelength to follow the thermal drift of the resonance position. For each pump power, the signal wavelength is scanned across the selected resonance at a detuning of $-10\;\text{nm}$. The peak CE is plotted versus coupled pump powers along with a theoretical model curve of slope $2$ in Fig. \ref{fig:CE_vs_power}. The experiment and the model agree well, and this means that nonlinear absorption can be neglected.

We show that AlGaAs resonators made of nanowaveguides exhibit enhanced nonlinearity with cw excitation of the fundamental TM mode. Nonetheless, as apparent from Fig. \ref{fig:CE_vs_dlambda}(c), we see a decreasing trend in the CE as the detuning from the pump wavelength increases, and this is due to the normal dispersion ($\beta^{(2)}>0$) of the TM mode, hence lacking a perfect phase matching (low $\eta$) and therefore a relatively limited FWM bandwidth.

On the other hand, we expect that the TE mode has a much broader FWM bandwidth because of the ZDW near $1550\;\text{nm}$ and an anomalous-dispersion regime ($\beta^{(2)}<0$) at longer wavelengths. Efficient FWM can lead to parametric oscillation. In a pump-only excitation, spontaneous four-wave mixing (SFWM) seeds the signal and idler photons into the ring cavity and with parametric gain larger than loss these photons build up, oscillate and possibly lead to cascaded FWM, culminating in a frequency comb. A broad FWM bandwidth would therefore allow octave-spanning frequency comb generation \cite{Okawachi2011}. In order to achieve this, the resonator requires a low round-trip loss and appropriate coupling with bus waveguides. A low-loss AlGaAs nanowaveguide is already achievable \cite{Apiratikul2014a} today whereas the coupler can be optimized via design and fabrication.

%%%\section{Conclusion}
In conclusion, we demonstrated low-power ($24\;\text{mW}$) cw-FWM wavelength conversion in an AlGaAs-nanowaveguide microresonator with a CE enhancement of $12\;\text{dB}$ and a maximum CE of $-43\;\text{dB}$. The experimental results match well with the theory which also predicts that the microresonator could potentially boosts the CE to $-15\;\text{dB}$. The device exhibits high ultrafast nonlinearity without detrimental effects from nonlinear absorption in the employed power range. Our result presents progress toward fully-integrated nonlinear FWM-based devices based on the versatile AlGaAs system, in particular for new frequency generation.

\end{document}